# Tunable subwavelength strong absorption by graphene wrapped dielectric particles

Bing Yang[1,2], Tong Wu[1] ,Yue Yang[1] , and Xiangdong Zhang[1]

[1]School of Physics, Beijing Institute of Technology, Beijing, 100081, China

[2]School of Physical Science and Information Engineering, Liaocheng University, 252059, Shandong, China;
Shandong Provincial Key Laboratory of Optical Communication Science and Technology, 252059, Shandong, China

E-mail: zhangxd@bit.edu.cn

**Abstract**

The optical absorption properties of graphene wrapped dielectric particles have been investigated by using Mie scattering theory and exact multi-scattering method. It is shown that subwavelength strong absorption in infrared spectra can take place in such systems due to the excitation of plasmon resonance in graphene. The absorption characteristics and efficiency are tunable by varying Fermi level and damping constant of graphene, or by changing size and dielectric constant of small particles. For a cluster of these particles, the absorption characteristics are also affected by the separation distance between them. These extreme light resonances and absorptions in graphene wrapped nanostructures have great potential for opto-electronic devices.

(Some figures may appear in colour only in the online journal)

## 1. Introduction

Graphene, with a two-dimensional (2D) form of carbon in which atoms are arranged in a honeycomb lattice is becoming a research focus in many fields recent years because of its unique electric, magnetic and optical properties due to its particular electronic band structure [1]. In opto-electronics aspects, there also has been a great deal of interest in studying the optical properties of graphene due to its abundant potential applications within a wide spectral range from terahertz (THz) to visible frequencies [2-12]. As an ultra-thin two-dimensional carbon material, graphene is widely used in the transparent electrodes and optical display materials [3–5], it has also been applied for opto-electronics components such as photodetectors [6-8], optical modulators [9, 10], and so on [11, 12]. In these applications, the strength of interaction between graphene and electromagnetic (EM) waves plays a central role. However, a single sheet of homogeneous graphene absorbs only about 2.3% of normal incidence light in the visible and near-infrared range [13]. The weak optical absorptions block the potential of graphene for opto-electronics applications in some sense. Thus, various methods to improve the interaction between graphene and EM waves have been proposed. To the best of our knowledge, there are mainly three kinds of strategies developed in recent years for this purpose. The first one is to combine graphene layer with conventional plasmonic nanostructures [14-17]. The plasmonic resonance of these conventional nanostructures, such as gold nanodisks, confines and enhances the electromagnetic radiation and hence improves the interaction between graphene and EM waves. The second method is to place graphene in different kinds of optical microcavites [7, 8, 18, 19]. The localization and enhancement of EM waves by the microcavities also can strengthen the interaction between graphene and EM waves. The third one is to fabricate periodically patterned graphene islands [9, 20-24]. In this case, graphene islands can support plasmon oscillations which can confine electromagnetic radiation, and through these plasmonic resonance graphene islands arrays can improve the interaction between graphene and EM waves and hence enhance the light absorptions. Because the charge carriers of patterned graphene can be easily tuned by electrical gating or chemical doping, tunable light absorption can be easily realized in this case. To date, many kinds of graphene island shapes have been proposed experimentally and theoretically, such as graphene ribbons, nanodisks, nanorings, nanotubes and so on [9, 23].

In this work, we propose a method by wrapping small dielectric particles with graphene to improve interaction between graphene and EM waves. By using the Mie scattering theory and exact multi-scattering method, we theoretically study the absorption characteristics of these particles. Results show that subwavelength strong absorption in infrared spectra can take place in these systems, and these kinds of absorptions are tunable though changing Fermi level and damping constant of graphene, or varying size and dielectric constant of small particles. For a cluster of these particles, the absorptions are also affected by the separation distance between them.

## 2. Theory and Method

We consider a cluster with $N$ monolayer-graphene wrapped dielectric spheres embedded in a background medium with relative electric permittivity $\varepsilon_b$ and relative magnetic permeability $\mu_b$. The spheres have radii $a$, relative permittivity $\varepsilon$ and relative permeability $\mu$. The geometry of the problem is shown in figure 1(b). It is worthy to point out that similar systems have been successfully fabricated recently where they used graphene wrapped gold nanoparticles for Raman imaging [25]. When the system we consider is irradiated by an external electromagnetic field with wavevector $\mathbf{k}$, the scattering field can be obtained by means of Mie scattering theory and multi-scattering method as briefly described below.

As graphene layer is an ultra-thin two-dimensional material only with one atom thickness ($\approx 0.3\text{nm}$)

in the radial direction on the sphere surface, the effect of graphene on the scattering properties of the graphene wrapped dielectric sphere can be attributed to the interfacial effect as shown in figure 1(a), which is reflected by a boundary condition in Mie scattering theory. For example, the magnetic field boundary condition for a monolayer graphene wrapped dielectric sphere can be expressed as

$$\hat{\mathbf{n}} \times (\mathbf{H}_0 + \mathbf{H}_s - \mathbf{H}_i) = \mathbf{J}, \tag{1}$$

where $\hat{\mathbf{n}}$ is the outwardly directed unit normal to the particle surface, $\mathbf{H}_0$, $\mathbf{H}_s$ and $\mathbf{H}_i$ represent the incident, scattered and internal magnetic fields of the dielectric sphere, respectively. $\mathbf{J} = \sigma_s \mathbf{E}_t$ is induced by the tangential component $\mathbf{E}_t$ of the electric field and is proportional to the optical conductivity $\sigma_s$ of graphene layer which can be derived from Kubo formula [26-30]. For illustration and simplicity, we only consider the situation where the absolute value of Fermi level $E_f$ of graphene is much larger than $k_B T$, where $k_B$ is the Boltzmann constant and $T$ is absolute temperature (or for low temperature conditions) [27, 30]. So we chose $\sigma_s = \sigma_d + \sigma_i$ with $\sigma_d$ the Drude term describing intra-band processes and $\sigma_i = \sigma_i' + i\sigma_i''$ describing inter-band contributions with

$$\sigma_d = \sigma_0 \frac{4E_f}{\pi(\Gamma_c - i\hbar\omega)}, \tag{2}$$

$$\sigma_i' = \sigma_0 \left(1 + \frac{1}{\pi} \arctan \frac{\hbar\omega - 2E_f}{\Gamma_c} - \frac{1}{\pi} \arctan \frac{\hbar\omega + 2E_f}{\Gamma_c}\right), \tag{3}$$

$$\sigma_i'' = -\sigma_0 \frac{1}{2\pi} \ln \frac{(2E_f + \hbar\omega)^2 + \Gamma_c^2}{(2E_f - \hbar\omega)^2 + \Gamma_c^2}. \tag{4}$$

Here $\sigma_0 = e^2/4\hbar$ is called the universal conductivity of graphene with the electronic charge $e$ and the reduced Plank constant $\hbar$, $\Gamma_c$ is the damping constant and $\omega$ is angular frequency of the incident field. The optical conductivity is the function of the incident wave frequency and Fermi level $E_f$ of graphene as $\Gamma_c$ is a constant. Such a scattering problem is similar to the charged sphere, which has been discussed in some references [31-33]. Together with the electric field boundary condition in Mie scattering theory $\hat{\mathbf{n}} \times (\mathbf{E}_0 + \mathbf{E}_s - \mathbf{E}_i) = 0$ ($\mathbf{E}_0$, $\mathbf{E}_s$ and $\mathbf{E}_i$ represent the incident, scattered and internal electric fields of the dielectric sphere, respectively). The coefficients of the scattering field of the single graphene wrapped sphere can be expressed as

$$a_{mn}^s = \frac{\psi_n(x)\psi_n'(mx) - m\psi_n'(x)\psi_n(mx) - i\tau\psi_n'(mx)\psi_n'(x)}{\psi_n'(mx)\xi_n(x) - m\psi_n(mx)\xi_n'(x) - i\tau\psi_n'(mx)\xi_n'(x)} a_{mn}^0, \tag{5}$$

$$b_{mn}^s = \frac{\psi_n(mx)\psi_n'(x) - m\psi_n'(mx)\psi_n(x) + i\tau\psi_n(mx)\psi_n(x)}{\psi_n(mx)\xi_n'(x) - m\psi_n'(mx)\xi_n(x) + i\tau\psi_n(mx)\xi_n(x)} b_{mn}^0, \tag{6}$$

where $a_{mn}^0$, $b_{mn}^0$ are expanding coefficients of vector spherical wave functions for the incident fields of the sphere located at position $\mathbf{r}_0$ ($x_0$, $y_0$, $z_0$), $\psi_n(x) = x j_n(x)$ and $\xi_n(x) = x h_n^{(1)}(x)$ are Riccati-Bessel functions, $j_n(x)$ and $h_n^{(1)}(x)$ are the sphere Bessel and Hankel functions of the first kind, $x = ka$ is the dimensionless size parameter with the sphere radius $a$ and wave number $k = (\varepsilon_b \mu_b)^{1/2} \omega/c$ with $c$ the light speed in vacuum. The prime indicates differentiation with respect to the argument in parentheses, $m = (\varepsilon \mu)^{1/2}/(\varepsilon_b \mu_b)^{1/2}$ is the complex relative refractive index of the sphere to the outside medium ($\varepsilon_b = 1$, $\mu_b = \mu = 1$ in our calculation). The dimensionless value $\tau = \sigma_s (\mu_0/\varepsilon_0)^{1/2}$ is defined as surface conductivity of the graphene wrapped sphere with $\mu_0$ and $\varepsilon_0$ the permeability and permittivity in vacuum respectively.

After the scattering problem from a scatter has been solved, the total scattering coefficients of a cluster of sphere particles shown in figure 1(b), $a_{mn}^{total}$ and $b_{mn}^{total}$, can be obtained by the multi-scattering method. The method has been introduced in detail in reference [34]. Thus, the extinction and scattering efficiencies, $Q_{ex}$ and $Q_{sc}$, can be expressed as

$$Q_{ex} = \frac{4}{Nx^2} \sum_{n=1}^{\infty} \sum_{m=-n}^{n} n(n+1)(2n+1) \frac{(n-m)!}{(n+m)!} \mathrm{Re}\left(a_{mn}^{0*} a_{mn}^{total} + b_{mn}^{0*} b_{mn}^{total}\right) , \qquad (7)$$

$$Q_{sc} = \frac{4}{Nx^2} \sum_{n=1}^{\infty} \sum_{m=-n}^{n} n(n+1)(2n+1) \frac{(n-m)!}{(n+m)!} \left(\left|a_{mn}^{total}\right|^2 + \left|b_{mn}^{total}\right|^2\right) , \qquad (8)$$

where $a_{mn}^{0*}$ and $b_{mn}^{0*}$ represent the complex conjugate of $a_{mn}^0$ and $b_{mn}^0$, respectively. Then, the absorption efficiency, $Q_{ab} = Q_{ex} - Q_{sc}$, can be obtained.

## 3. Numerical Results and Discussion

We first consider the absorption efficiency of the single graphene wrapped dielectric sphere illuminated by the linear polarization incident wave. Figure 2 describes the calculated results of $Q_{ab}$ as a function of wavelength for different tunable variables of the system. The strong resonance absorption in the infrared region is found, which depends on the system size, external electromagnetic fields and dielectric constants of spheres. This is because such a strong absorption comes from the plasmon resonance in graphene with the external EM waves. Due to the peculiar electronic band structure, the optical conductivity is dominated by the intra-band term expressed by equation (2) in high doped graphene

($E_f > \hbar\omega$), leading to graphene acts like metals in our interested frequencies. Therefore, such a resonance absorption strongly depends on the size of the particle. With the decrease of size, the resonance wavelength becomes small, blue shift of the resonance peak occurs, full width at half maximum of the peak becomes small and the amplitude is improved as shown in figure 2(a). For example, the resonance absorption peak appears at wavelength 67 $\mu$m for the dielectric sphere with $a = 5$ $\mu$m and $\varepsilon = 3.9$. It appears at wavelength 8.91 $\mu$m for the dielectric sphere with $a = 100$ nm, the resonance wavelength is about one hundred times as large as the radius of the sphere in such a case. Thus, it is typical subwavelength behavior.

Similarly, if we change the Fermi level of graphene, which can be realized by tuning the density of charge carriers through the external electrical gating field and/or chemical doping, the surface plasmon resonance also changes. The blue shift of the resonance peak also occurs with the increase of $E_f$ from 0.3 eV to 1.0 eV as shown in figure 2(b). Our results qualitatively coincide with the recent research conclusions of other groups [9, 23] for graphene ribbons and graphene nanotubes. It is interesting that similar phenomenon can be found if we only change the dielectric constant inside the sphere as shown in figure 2(c). Blue shift and enhancement of absorption efficiencies also appear as the dielectric constant of sphere decreases from 12.96 to 3.9. This is because the decrease of dielectric constant inside the sphere is equivalent to the decrease of size of the system. Of course, the damping constant $\Gamma_c$ of graphene is also a tunable parameter. With the increase of $\Gamma_c$ from 0.2 meV to 1.5 meV, the resonance absorption becomes small and broadens in absorption wavelength range due to the more momentum randomization of electrons in graphene as shown in figure 2(d).

In contrast to the case of the single graphene wrapped dielectric sphere, the absorption spectra of a cluster with many graphene wrapped dielectric spheres exhibit richer phenomena. Figure 3(a) and (b) show the absorption efficiency of two-sphere systems at the incident direction k of the linear polarized wave perpendicular or parallel to the longitudinal axis of the system, respectively (see the insets in the corresponding subplots, the polarization of the incident wave is denoted by red arrow with E). Here the radius and dielectric constant of two spheres are taken as $a = 100$ nm and $\varepsilon = 12.96$, parameters of graphene are taken as $E_f = 1.0$ eV and $\Gamma_c = 1.0$ meV. The grass-green, red, cyan, magenta and yellow lines represent the separation distance between two spheres $d = 5$ nm, 20 nm, 50 nm, 100 nm, 300 nm, respectively. For comparison, the case of the single sphere is plotted in blue line. When the separation distance is large enough, the absorption phenomenon is identical with the case of the single sphere. With the decrease of the separation distance, red shift of the resonance absorption peak occurs at the incident direction of wave perpendicular to the longitudinal axis of the system (see figure 3(a)). In contrast, blue shift appears at the incident direction of wave parallel to the longitudinal axis of the system (see figure 3(b)). This is because the coupling effect between the two particles becomes strong with the decrease of the separation distance. In comparison with the single sphere, the effective transverse size along the polarization direction of incident wave becomes large for the case at the incident direction of wave perpendicular to the longitudinal axis (figure 3(a)), the resonance absorption peak shifts to the long wavelength. For the case at the incident direction of wave parallel to the longitudinal axis of the system (figure 3(b)), the effective longitudinal size become large, which is equivalent to decrease of the effective transverse size, the resonance absorption peak shifts to the short wavelength.

The absorption efficiency not only depends on the parameters of scatters, it is also related to the incident wave. The above discussions only focus on the linear polarized incident wave. Figure 3(c) shows the absorption efficiency of two-sphere systems at the circular polarization incident wave with k perpendicular to the longitudinal axis of the system (see inset of figure 3(c)). Here the parameters of scatters are taken identical with those in figure 3(a). Because the circular polarized incident wave includes two kinds of polarized information, parallel and perpendicular to the longitudinal axis of the system, the absorption spectra in figure 3(c) exhibit double resonant absorption peaks, which are caused by the two kinds of the coupling resonances as in cases of figure 3(a) and (b), respectively. Due to the subwavelength behavior, these resonant absorption peaks are mainly determined by the dipole term in the multi-scattering calculation. In figure 3(a), the dipoles induced in two spheres are end to end and perform a strong interactions because of the short distance between them (equal to the separation distance between two spheres), while in figure 3(b), the dipoles induced in two spheres are side by side and perform a weak interactions due to the long distance between them (equal to the distance of two spheres centers). Hence the blue shift of absorptions in figure 3(b) is much less than the red shift of absorptions in figure 3(a). The same results also can be found in figure 3(c). Although the dipolar effects domain the resonance absorptions, the multipole effects can also exhibit in the absorption spectra. In the inset of figure 3(c), the multipole signal is enlarged. We find that it is sticking out with the decrease of the separate distance between two spheres.

For the cluster with many graphene wrapped dielectric spheres, we choose two representative cases as shown in the insets of figure 4(a) and (b). For the first case, 7 spheres are arranged in a plane with six spheres at the hexagon vertex and the seventh at the center of the hexagon and the second case is 27 spheres are arranged in three-dimensional (3D) space with a stacking manner of $3 \times 3 \times 3$. The direction of incident electromagnetic waves is perpendicular to the sphere layer plane with linear polarization along the longitudinal axis of two adjacent spheres. For comparison, the absorption efficiency of the single sphere is also plotted in blue line. The parameters of spheres and graphene are taken identical with those in figure 3(a). From figure 4 we can see resonance absorption peaks exhibit red shift as separation distances between spheres decrease for both conditions. Compared with the 7 spheres system, the 27 spheres system shows relatively less shifts for the same separation distance of spheres. As mentioned above, for scattering of these subwavelength structures, dipole approximation for plasmon resonance can be employed to qualitatively understand the wavelength shift of the resonance absorptions. For the case of 7 spheres system, interactions of dipoles in adjacent sphere layers parallel to the E-k plane compete with the interactions of dipoles end to end and hence lead to less redshift for the same separation distance of spheres compared with two spheres case in figure 3(a). For the case of 27 spheres system, except for the similar interactions of dipoles in 7 sphere system, dipoles in adjacent sphere layers perpendicular to the propagation direction of incident wave are side by side and hence the interactions of these dipoles further reduce the redshifts of the resonance absorption compared with 7 sphere system. At the same time, the number of resonance peaks also increases due to the complex coupling interactions of plasmon resonances.

In order to reveal the absorption efficiency of graphene wrapped dielectric spheres more clearly, we calculate the reflectance, transmittance and absorptance of light in a planar structure consisting of a square array of graphene wrapped spheres as shown in inset of figure 5 using the method in Ref. [35, 36]. The calculated results for the absorptance ($A$, denoted by solid lines), transmittance ($T$, denoted by dashed lines) and reflectance ($R$, denoted by cycle symbols) are plotted in figure 5. Here the parameters of scatters are taken identical with those in figure 3(a). $T$ and $R$ are defined as the ratio of the transmitted, respectively the reflected, energy flux to the energy flux associated with the incident wave. The absorptance is defined from the requirement of energy conservation $A = 1 - T - R$. The red, cyan, magenta and yellow lines correspond to the case with the separation distance between two spheres $d$ =20 nm, 50 nm, 100 nm and 300 nm, respectively. Due to the interaction across the lattice, significant red shifts take place as separation distance of spheres decreases. It is shown clearly that the transmittance increase and reflectance decrease as the separation distances increase. The absorptance can reach 50% as the suitable lattice constants are taken ($d$ =50 nm and 100 nm in this case) showing the strong absorptions of incident infrared wave. It should be pointed out that because of the infinite extend of the planar structure of the spheres, the redshift of resonance absorption is more serious than that of two spheres case in figure 3(a) for the same separation distance between spheres.

## 4. Summary

In summary, we have studied the optical absorption properties of graphene wrapped dielectric particles based on the Mie scattering theory and exact multi-scattering method. We have demonstrated strong subwavelength absorption of graphene wrapped dielectric spheres due to the plasmon resonance of graphene with external electromagnetic waves. By tuning the Fermi level and damping of graphene, or changing dielectric constant and size of spheres, the absorption efficiency and resonance wavelength can be tuned easily. The blue and red shifts of resonant absorption peaks have been found in the system for the linear polarized wave with different incident directions. The multipole signal has also been observed in the absorption spectra. Our researches propose a new idea to realize tunable strong absorption of graphene in infrared spectra using subwavelength sphere configurations. The methods and results can be helpful in light modulation, solar cell, biological imaging and even in THz communication.

## Acknowledgments

This work was supported by the National Natural Science Foundation of China (Grant No. 11274042), the National Key Basic Research Special Foundation of China under Grant 2013CB632704 and Shandong Province Natural Science Foundation of China (No. ZR2012FL20).

Figure Captions

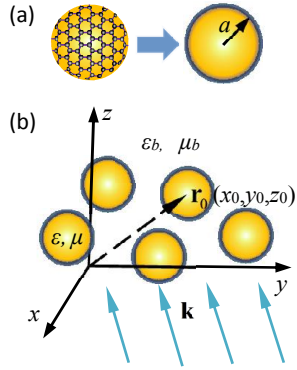

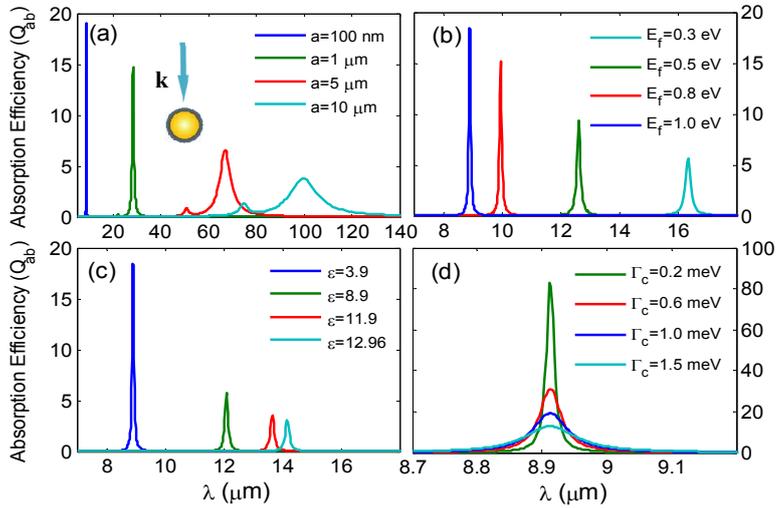

**Figure 1.** (a) Schematic diagram and model for a graphene wrapped dielectric sphere. (b) Geometry and coordinate of the scattering problem for a cluster of graphene wrapped dielectric spheres. The system is scattered by an incident plane wave with wavevector **k**.

**Figure 2.** Absorption efficiency of single graphene wrapped dielectric sphere as a function of wavelength under the excitation of linear polarization incident wave. (a) Various sizes of the sphere at $E_f$ =1.0 eV, $\Gamma_c$ =1.0 meV and $\varepsilon$ =3.9. (b) Different $E_f$ at $a$ =100 nm, $\varepsilon$ =3.9 and $\Gamma_c$ =1.0 meV. (c) Different dielectric constant at $a$ =100 nm, $E_f$ =1.0 eV and $\Gamma_c$ =1.0 meV. (d) Different $\Gamma_c$ at $a$ =100 nm, $\varepsilon$ =3.9 and $E_f$ =1.0 eV.

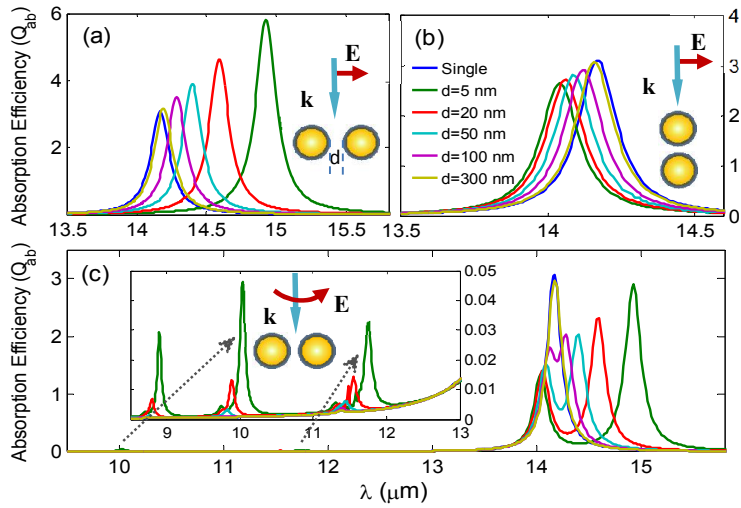

**Figure 3.** Absorption efficiency of two-graphene wrapped dielectric spheres as a function of wavelength under the excitation of linear and circular polarization incident waves. The radii and dielectric constant of spheres are taken as $a$ =100 nm and $\varepsilon$ =12.96. The parameters for monolayer graphene are taken as $E_f$ =1.0 eV and $\Gamma_c$ =1.0 meV. (a) is for perpendicular incidence with electric polarization **E** along the longitudinal axis of the system; (b) parallel incidence; (c) perpendicular incidence with circular polarization. The Inset shows the enlarged results for multipole effects.

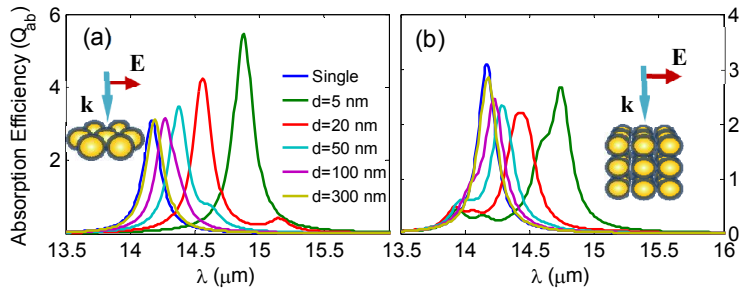

**Figure 4.** absorption efficiency of many graphene wrapped dielectric spheres as a function of wavelength under the excitation of the linear polarization incident wave. (a) 7-sphere system in a plane as shown in Inset; (b) 27-sphere system in 3×3×3 packing way as shown in Inset. Here the parameters of graphene wrapped dielectric spheres are taken identical with those in figure 3. The electric polarization **E** of the incident wave is along longitudinal axis of two adjacent spheres.

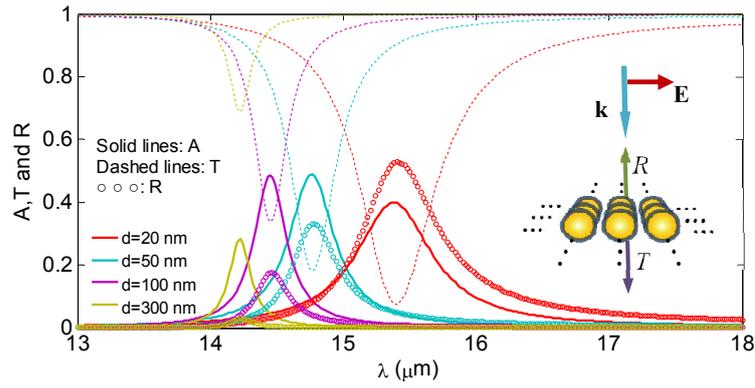

**Figure 5.** Absorptance (*A*, denoted by solid lines), transmittance (*T*, denoted by dashed lines) and reflectance (*R*, denoted by cycle symbols) of an infinite layer of graphene wrapped dielectric spheres arranged in square lattice as a function of wavelength under the excitation of the linear polarization incident wave at various separation distances. The parameters of scatters are taken identical with those in figure 3. The electric polarization **E** of the incident wave is along longitudinal axis of two adjacent spheres. The absorptance are defined from the requirement of energy conservation $A = 1 - T - R$.